# Failure of the work-Hamiltonian connection for free energy calculations

Jose M. G. Vilar[1] and J. Miguel Rubi[2]

[1]*Computational Biology Program, Memorial Sloan-Kettering Cancer Center, 1275 York Avenue, New York, NY 10021*

[2]*Departament de Fisica Fonamental, Universitat de Barcelona, Diagonal 647, 08028 Barcelona, Spain*

Abstract

Extensions of statistical mechanics are routinely being used to infer free energies from the work performed over single-molecule nonequilibrium trajectories. A key element of this approach is the ubiquitous expression $dW/dt = \partial H(x,t)/\partial t$, which connects the microscopic work $W$ performed by a time-dependent force on the coordinate $x$ with the corresponding Hamiltonian $H(x,t)$ at time $t$. Here we show that this connection, as pivotal as it is, cannot be used to estimate free energy changes. We discuss the implications of this result for single-molecule experiments and atomistic molecular simulations and point out possible avenues to overcome these limitations.

Hamiltonians provide two key ingredients to bridge the microscopic structure of nature with macroscopic thermodynamic properties: they completely specify the underlying dynamics and they can be identified with the energy of the system [1]. At equilibrium, the link with the thermodynamic properties is established through the partition function $Z = \int e^{-\beta H(x)} dx$, which here uses the Hamiltonian $H(x)$ in the coordinate space $x$ as the energy of the system [2]. In particular, the free energy is given by $G = -\frac{1}{\beta} \ln Z$, where $\beta \equiv 1/k_B T$ is the inverse of the temperature $T$ times the Boltzmann's constant $k_B$. Thermodynamic properties play an important role because they provide information that is not readily available from the microscopic properties, such as whether or not a given process happens spontaneously.

The connection between work and Hamiltonian expressed through the relation $\frac{d}{dt} W = \frac{\partial}{\partial t} H(x,t)$, or equivalently through its integral representation $W = \int_0^t \frac{\partial}{\partial t'} H(x(t'),t') dt'$, is typically used to extend statistical mechanics to far-from-equilibrium situations [3-5]. These relations are meant to imply that the work $W$ performed on a system is used to change its energy. The potential advantage of this type of approach is that it would allow one to infer thermodynamic properties even when the relevant details of the Hamiltonian are not known or when they are too complex for a direct analysis. Experiments and computer simulations can thus be performed to probe the microscopic mechanical properties from which to obtain thermodynamic properties. Time-dependent Hamiltonians, however, provide the energy up to an arbitrary factor that typically depends on time and on the microscopic history of the system. Such dependence, as we show below, prevents this approach from being generally applicable to compute thermodynamic properties.

To illustrate how work and Hamiltonian fail to be generally connected, we consider a system described by the Hamiltonian $H_0(x)$ under the effects of a time-dependent force $f(t)$. The total Hamiltonian is given by

$$H(x,t) = H_0(x) - f(t)x + g(t),$$

where $g(t)$ is an arbitrary function of time, which leads to a total force $F = -\partial H_0/\partial x + f(t)$. The function $g(t)$ does not affect the total force but it changes the Hamiltonian. Therefore, $g(t)$ has to be chosen so that the Hamiltonian can be identified with the energy of the system.

In general, the arbitrary time dependence of the Hamiltonian, $g(t)$, cannot be chosen so that the Hamiltonian gives a consistent energy. Consider, for instance, that the system, being initially at $x_0$, is subjected to a sudden perturbation $f(t) \equiv f_0 \Theta(t)$, where $f_0$ is a constant and $\Theta(t)$ is the Heaviside step function. The work performed on the system, $W = f_0(x_t - x_0)$, where $x_t \equiv x(t)$ represents the value of the coordinate $x$ at time $t$, is in general different from $\int_0^t \frac{\partial}{\partial t'} H(x_{t'}, t') dt' = -f_0 x_0 + g(t) - g(0)$, irrespective of the explicit form of the function $g(t)$.

To illustrate the consequences of the lack of connection between work and changes in the Hamiltonian, we focus on the domain of validity of nonequilibrium work relations [3] of the type

$$e^{-\beta \Delta G_E} = \left\langle e^{-\beta W} \right\rangle,$$

which have been widely used recently to obtain estimates $\Delta G_E$ of free energy changes from single-molecule pulling experiments [6] and atomistic computer simulations [7]. The promise of this type of relations is that they provide the values of the free energy from irreversible trajectories and therefore do not require equilibration of the system. Yet, in almost all instances in which this approach has been applied, the agreement with the canonical thermodynamic results has not been complete and in some cases the discrepancies have been large. These discrepancies have been attributed to the presence of statistical errors in the estimation of the exponential average $\langle e^{-\beta W}\rangle$ [8].

Currently, the mathematical validity of these type of nonequilibrium work relations appears to be well established: they have been derived using approximations [3] and rigorously for systems described by Langevin equations [4, 5]. However, all these derivations rely in different ways on the work-Hamiltonian connection, which as we show below prevents them from giving general estimates of thermodynamic free energies.

The free energy difference between two states is defined as $\Delta G = \langle W_{rev}\rangle$, where $W_{rev}$ is the work required to bring the system from the initial to the final state in a reversible manner [2]. Note that, if the system is not macroscopic, $W_{rev}$ is in general a fluctuating quantity. At quasi-equilibrium, the external force $f(t)$ balances with the system force $-\partial H(x)/\partial x$. After integration by the displacement, the reversible work done on the system is given by $W_{rev} = H_0(x_t) - H_0(x_0)$. Therefore, the free energy follows from

$$\Delta G = \iint W_{rev} P_{eq}(x_t,t) P_{eq}(x_0,0) dx_t dx_0 ,$$

where the equilibrium probabilities $P_{eq}$ are obtained, in the usual way, from the Boltzmann distribution $P_{eq}(x,t) = \frac{1}{Z(t)} e^{-\beta H(x,t)}$. To be explicit, let us consider a harmonic system described by $H_0(x) = \frac{1}{2}kx^2$ and $g(t) = 0$, with $k$ a constant. In this case, we can compute exactly the free energy change:

$$\Delta G = \frac{1}{2} k x_{eq}^2,$$

where $x_{eq} \equiv f(t)/k$, which leads to a positive value as required for non-spontaneous processes.

One might have been tempted to use the partition function to estimate changes in free energy according to the expression $\Delta G_Z = -\frac{1}{\beta} \ln(Z(t)/Z(0))$, where $Z(t) = \int e^{-\beta H(x,t)} dx$ is the time-dependent quasi-equilibrium partition function [3, 4]. However, this relation is not valid when changes in the Hamiltonian cannot be associated with changes in energy. In the case of the harmonic potential, the use of the time-dependent partition function leads to $\Delta G_Z = -\frac{1}{2} k x_{eq}^2$, a negative value inconsistent with a process that is not spontaneous. More generally, the Hamiltonian $H(x,t) = \frac{1}{2} k x_t^2 - f(t)(x-\gamma)$, where $\gamma$ is a constant parameter that does not affect the dynamics of the system, leads to $\Delta G_Z = k x_{eq}(\gamma - \frac{1}{2} x_{eq})$, which can be positive or negative depending on the value of $\gamma$. Therefore, the estimates $\Delta G_Z$ are not suitable to predict typical thermodynamic properties, such as whether or not a process happens spontaneously.

To what extent does the failure of the work-Hamiltonian connection impact nonequilibrium work equalities? In the case of a sudden perturbation and a harmonic potential discussed previously, the following result follows straightforwardly:

$$\langle e^{-\beta W}\rangle = \iint e^{-\beta f_0(x_t-x_0)} P_{eq}(x_t,t)P_{eq}(x_0,0)dx_t dx_0 = 1,$$

which is different from $e^{-\beta \Delta G}$.

An intriguing question then arises: why do experiments and computer simulations sometimes lead to results that agree with nonequilibrium work equalities? Let us consider a situation closer to the experimental and computational setups, with a harmonic time-dependent force that constrains the motion on the coordinate $x$:

$$H(x,t) = H_0(x) + \tfrac{1}{2}K(x-X_t)^2.$$

Here $K$ is a constant and $X_t$ is the time-dependent equilibrium position for the constraining force. In this case, with $H_0(x) = \tfrac{1}{2}kx^2$ and $X_0 = 0$, we also have

$$\Delta G = \langle W_{rev}\rangle = \frac{1}{2}kx_{eq}^2,$$

where now $x_{eq} \equiv \frac{K}{k+K}X_t$.

For quasi-equilibrium displacements of $X_t$, so that the work performed is equal to the reversible work, $W = W_{rev} = H_0(x_t) - H_0(x_0)$, we have

$$\langle e^{-\beta W_{rev}}\rangle = \iint e^{-\beta(H_0(x_t)-H_0(x_0))} P_{eq}(x_t,t)P_{eq}(x_0,0)dx_t dx_0,$$

which leads to

$$\langle e^{-\beta W_{rev}}\rangle = \frac{e^{-\beta \frac{k(k+K)}{2(2k+K)}x_{eq}^2}(k+K)}{\sqrt{K(2k+K)}}.$$

This result indicates that quasi-equilibrium does not guarantee the accuracy of the exponential estimate of the free energy from nonequilibrium work relations. The free energy change $\Delta G$ and its exponential estimate $\Delta G_E$ agree with each other only for large values of $K$. The reason is that, in this case, work and Hamiltonian are connected to each other when both quasi-equilibrium and large-$K$ conditions are fulfilled simultaneously. Under such conditions, the work-Hamiltonian connection is valid because $x \approx x_{eq} \approx X_t$ implies that the rate of change of the Hamiltonian, $\partial H(x,t)/\partial t = -K(x - X_t)dX_t/dt$, equals the power associated with the external force, $dW/dt = -K(x - X_t)dx/dt$. Interestingly, large values of $K$ suppress fluctuations and lead to quasi-deterministic dynamics. Indeed, the experimental data [6] and computer simulations [7] indicate that the agreement between the free energy change $\Delta G$ and its exponential estimate $\Delta G_E$ occurs mainly for relatively slow perturbations that lead to quasi-deterministic trajectories.

Bringing thermodynamics to nonequilibrium microscopic processes [9] is becoming increasingly important with the advent of new experimental and computational techniques able to probe the properties of single molecules [6, 7]. Our results show that the classical connection between work and changes in the Hamiltonian cannot be applied straightforwardly to time-dependent systems. As a result, quantities that are based on the work-Hamiltonian connection, such as those obtained from nonequilibrium work relations and time-dependent partition functions, cannot generally be used to estimate thermodynamically consistent free energy changes. A possible avenue to overcome these limitations, as we have shown here, is to identify the particular conditions for which work and changes in the Hamiltonian are connected to each other.